\documentclass[epj]{webofc}
\usepackage[utf8]{inputenc}
\usepackage[varg]{txfonts}   % Web of Conferences font
\usepackage{booktabs}
\usepackage{xcolor}
\definecolor{darkred}{rgb}{0.4,0.0,0.0}
\definecolor{darkgreen}{rgb}{0.0,0.4,0.0}
\definecolor{darkblue}{rgb}{0.0,0.0,0.4}
\usepackage[bookmarks,linktocpage,colorlinks,
    linkcolor = darkred,
    urlcolor  = darkblue,
    citecolor = darkgreen]{hyperref}
\wocname{EPJ Web of Conferences}
\woctitle{Lattice2017}
\begin{document}
%%%%%%%%%%%%%%%%%%%%%%%%%%%%%%%%%%%%%%%%%%%%%%%%%%%%%%%%%%%%%%%%%%%%%%%%%%%%%
%
\selectlanguage{english}
%----------------------------------------------------------------------------
\title{%
Multi-Grid Lanczos
}
%----------------------------------------------------------------------------
\author{%
\firstname{M.~A.} \lastname{Clark}\inst{1}\fnsep
\and
\firstname{Chulwoo} \lastname{Jung}\inst{2}\fnsep
\and
\firstname{Christoph} \lastname{Lehner}\inst{2}\fnsep\thanks{Speaker, \email{clehner@bnl.gov}}
% \and
%\firstname{Harry} \lastname{Potter}\inst{2}
}
%----------------------------------------------------------------------------
\institute{%
NVIDIA Corporation, 2701 San Tomas Expressway, Santa Clara, CA 95050, USA
\and
Physics Department, Brookhaven National Laboratory, Upton, NY 11973, USA
%Last address unknown
}
%----------------------------------------------------------------------------
\abstract{%
  We present a Lanczos algorithm utilizing multiple grids that reduces
  the memory requirements both on disk and in working memory by one
  order of magnitude for RBC/UKQCD's 48I and 64I ensembles at the
  physical pion mass. The precision of the resulting eigenvectors is
  on par with exact deflation.  
}
%----------------------------------------------------------------------------
\maketitle
%----------------------------------------------------------------------------
\section{Introduction}\label{intro}
In recent years RBC/UKQCD has benefited significantly from the
generation of the 2000 lowest eigenvectors of the preconditioned
normal (z)Mobius Domain Wall Fermion Dirac operator for the light
quarks on the 48I ($a^{-1}=1.7$ GeV) and 64I ($a^{-1}=2.3$ GeV)
ensembles at near physical pion mass. These eigenvectors were used
for deflation and volume averages over the low-mode space and were a
key ingredient in the on-going $g-2$ projects \cite{ref1,ref2,ref3};
they have also found additional use in the calculation of $\Delta M_K$
\cite{ref4}.

The storage cost for these vectors is substantial with 9.3 TB and 36
TB per configuration for the 48I and 64I ensembles
respectively. These high storage requirements both on disk and in RAM
are addressed in this contribution allowing for usage of these
methods at even larger volumes. Our approach makes deflation much more
applicable to architectures with limited amounts of high-bandwidth
memory such as GPUs and allows for running on small-scale clusters.

We note that related ideas were recently successfully used in the context
of Monte-Carlo estimation of the trace of a matrix inverse \cite{ref0}.

\section{Eigenvector compression}\label{sec-1}
\begin{figure}[tbp]
  \centering
  \includegraphics[width=9cm]{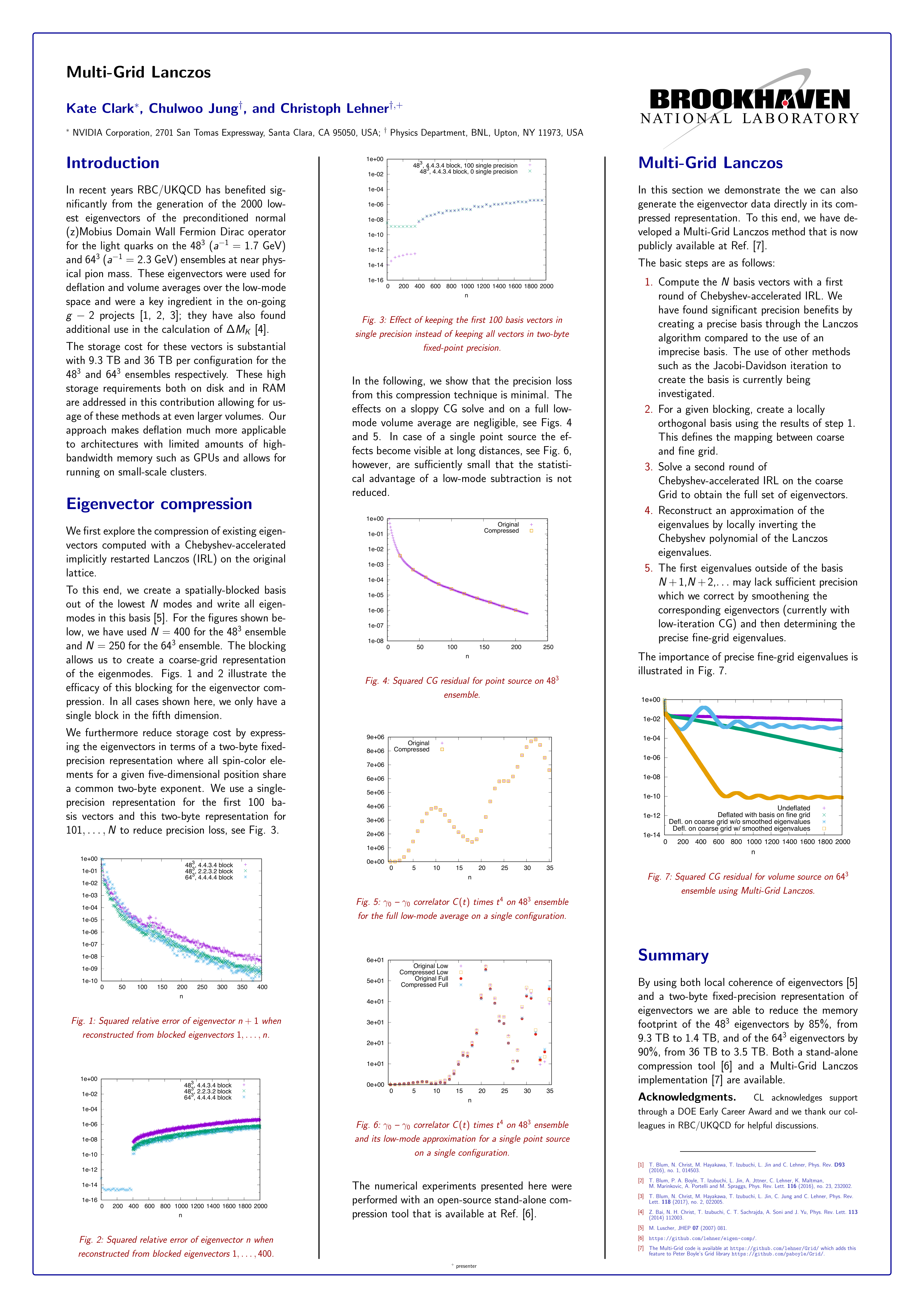}
  \caption{Squared relative error of eigenvector $n+1$ when reconstructed from blocked eigenvectors $1,\ldots,n$.}
  \label{fig1}
\end{figure}

\begin{figure}[tbp]
  \centering
  \includegraphics[width=9cm]{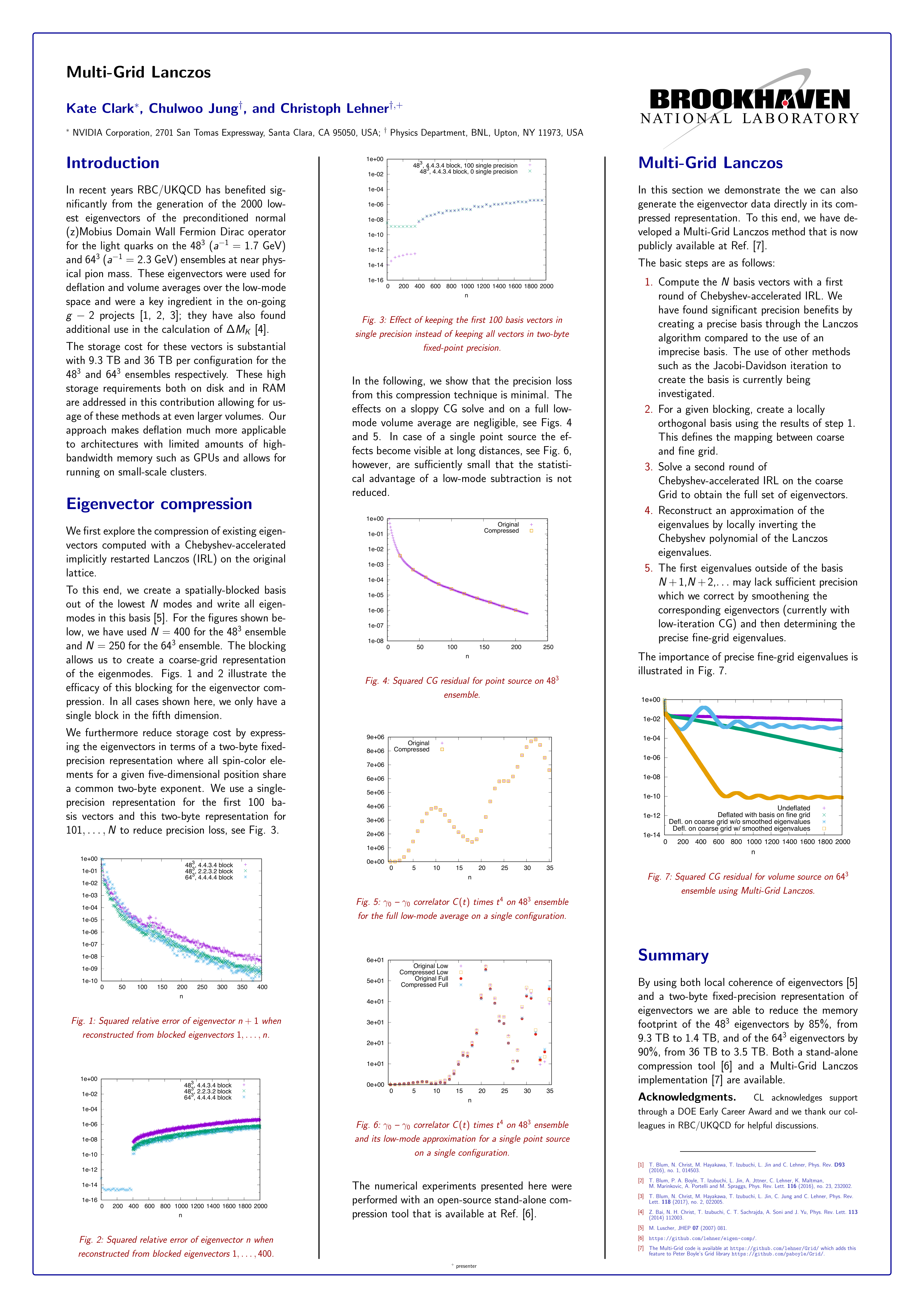}
  \caption{Squared relative error of eigenvector $n$ when reconstructed from blocked eigenvectors $1,\ldots,400$.}
  \label{fig2}
\end{figure}

We first explore the compression of existing eigenvectors computed
with a Chebyshev-accelerated implicitly restarted Lanczos (IRL) on the
original lattice.  To this end, we create a spatially-blocked basis
out of the lowest N modes and write all eigenmodes in this basis
\cite{ref5}.  For the figures shown below, we have used $N=400$ for
the 48I ensemble and $N=250$ for the 64I ensemble.  The blocking
allows us to create a coarse-grid representation of the eigenmodes.
Figs.~\ref{fig1} and \ref{fig2} illustrate the efficacy of this
blocking for the eigenvector compression.  The squared relative error
is the squared norm of the difference of original and reconstructed
vector divided by the squared norm of the original vector.
In all cases shown here, we
only have a single block in the fifth dimension.

\begin{figure}[tbp]
  \centering
  \includegraphics[width=9cm]{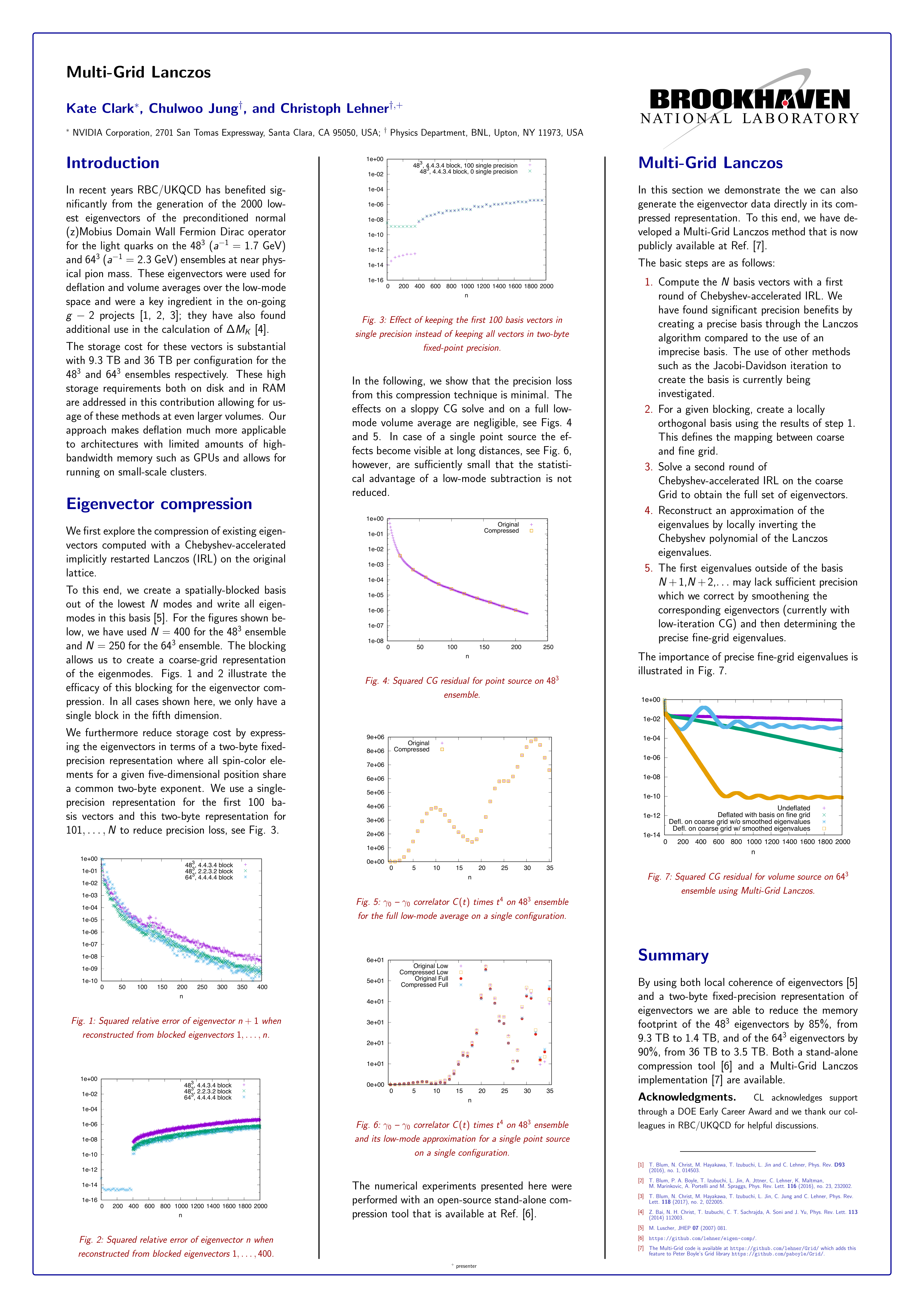}
  \caption{Effect of keeping the first 100 basis vectors in single precision instead of keeping all vectors in two-byte fixed-point precision.}
  \label{fig3}
\end{figure}

We furthermore reduce storage cost by expressing the eigenvectors in
terms of a two-byte fixed-precision representation, where all
spin-color elements for a given five-dimensional position share a
common two-byte exponent. We use a single-precision representation for
the first 100 basis vectors and this two-byte representation for
$101,\ldots,N$ to reduce precision loss, see Fig.~\ref{fig3}.

\begin{figure}[tbp]
  \centering
  \includegraphics[width=9cm]{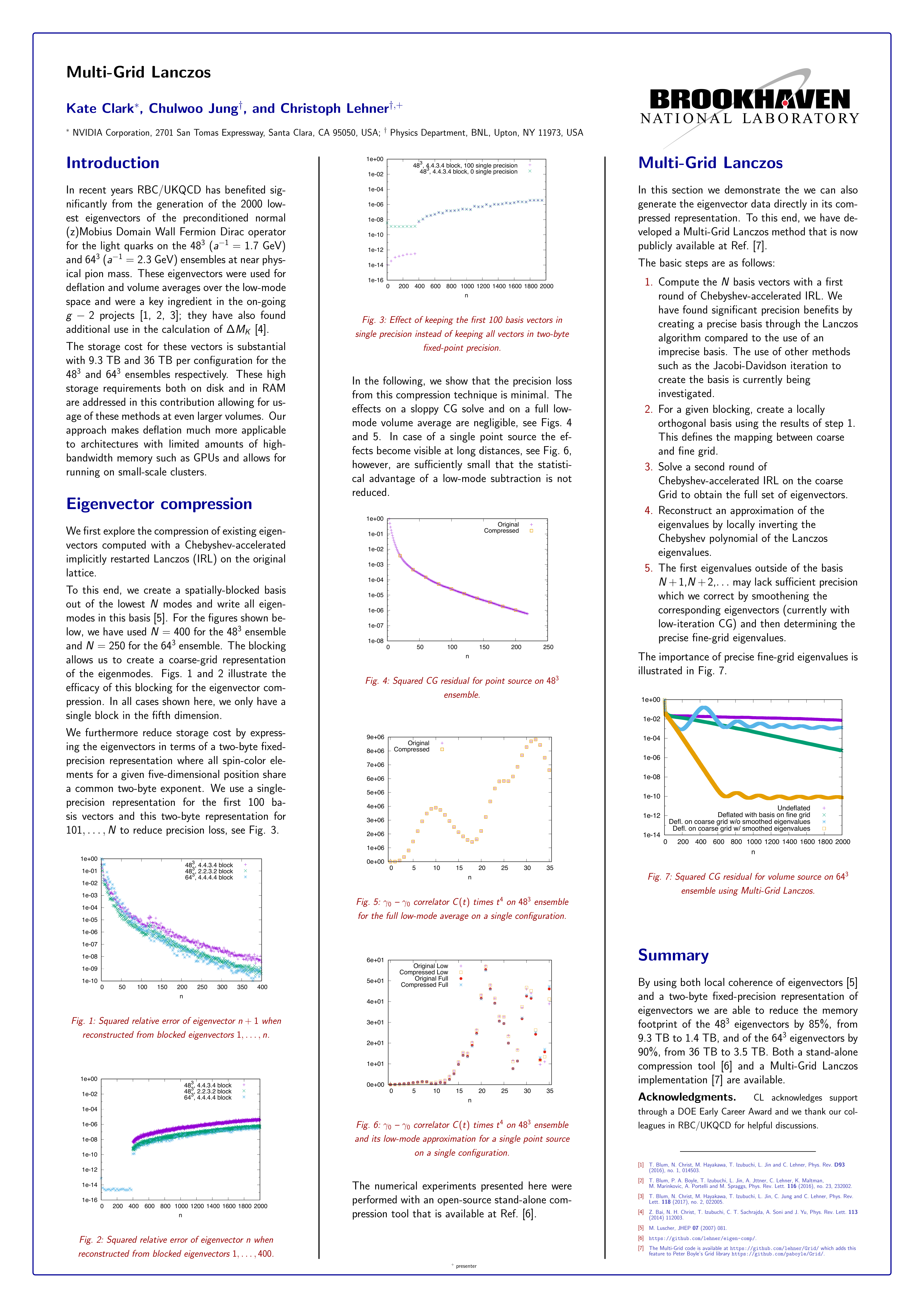}
  \caption{Squared CG residual as function of iteration number for point source on 48I ensemble.}
  \label{fig4}
\end{figure}

\begin{figure}[tbp]
  \centering
  \includegraphics[width=9cm]{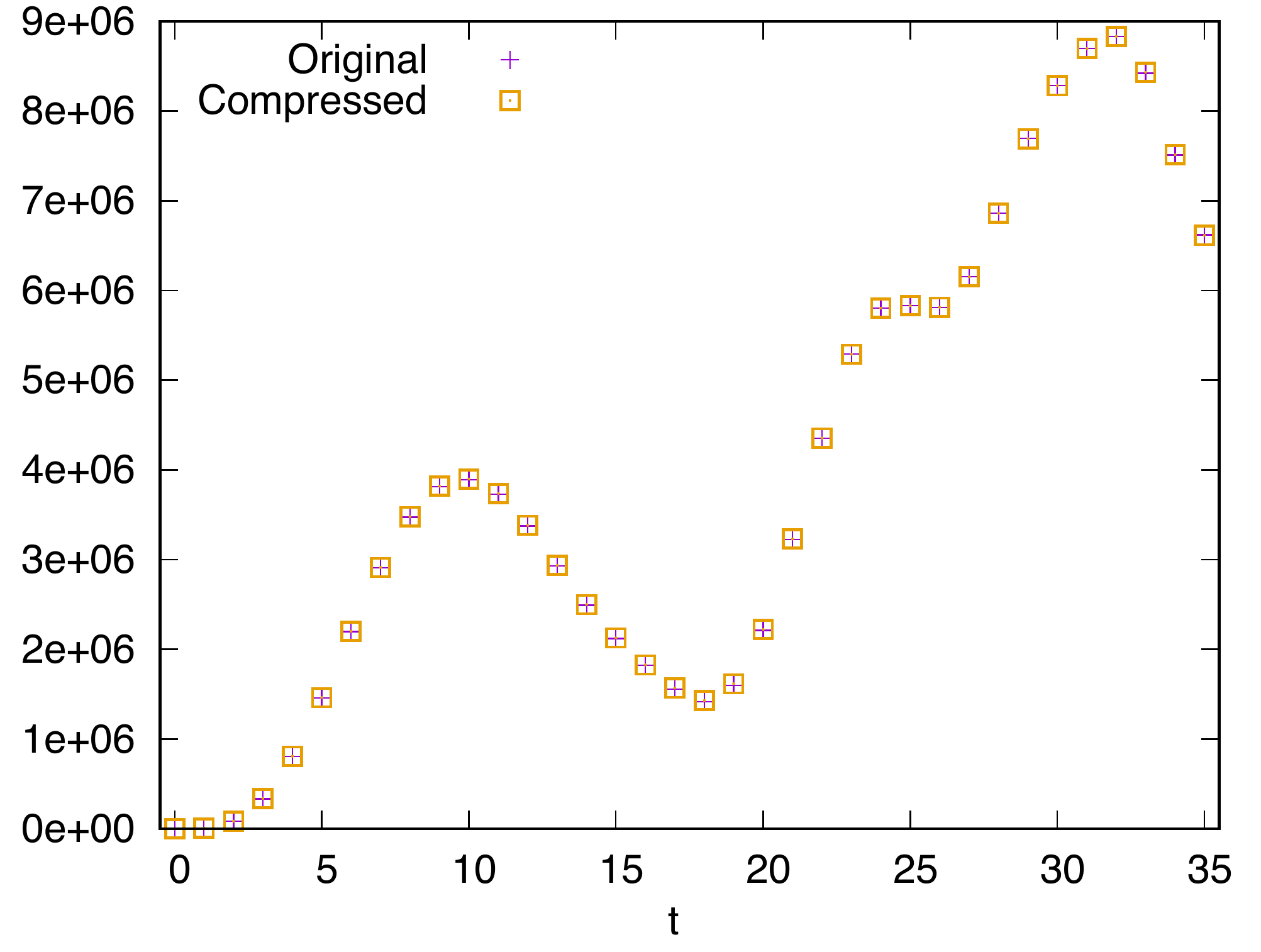}
  \caption{$\gamma_0$--$\gamma_0$ correlator $C(t)$ times $t^4$ on 48I ensemble for the full low-mode average on a single configuration.}
  \label{fig5}
\end{figure}

\begin{figure}[tbp]
  \centering
  \includegraphics[width=9cm]{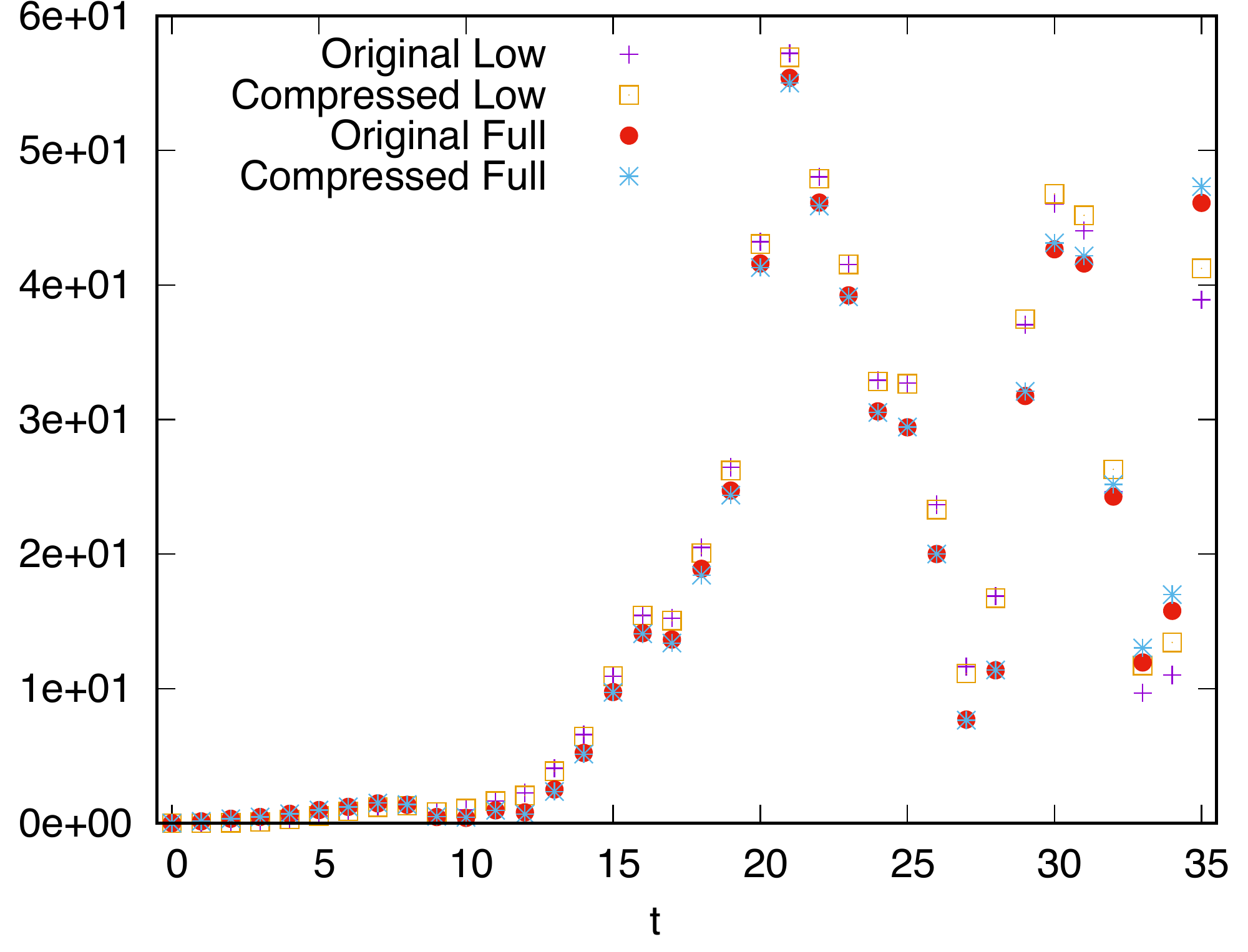}
  \caption{$\gamma_0$--$\gamma_0$ correlator $C(t)$ times $t^4$ on 48I ensemble and its low-mode approximation for a single point source on a single configuration.}
  \label{fig6}
\end{figure}

In the following, we show that the precision loss from this
compression technique is minimal. The effects on a sloppy CG solve and
on a full low-mode volume average are negligible, see Figs.~\ref{fig4}
and \ref{fig5}.  In case of a single point source the effects become
visible at long distances, see Fig.~\ref{fig6}, however, are
sufficiently small that the statistical advantage of a low-mode
subtraction is not reduced.

The numerical experiments presented here were performed with an
open-source stand-alone compression tool that is available at
Ref.~\cite{ref6}.

\section{Multi-Grid Lanczos}
In this section we demonstrate that we can also generate the
eigenvector data directly in its compressed representation. To this
end, we have developed a Multi-Grid Lanczos method that is now
publicly available at Ref.~\cite{ref7}.

The basic steps are as follows:

\begin{enumerate}
\item Compute the N basis vectors with a first round of
  Chebyshev-accelerated IRL. We have found significant precision
  benefits by creating a precise basis through the Lanczos algorithm
  compared to the use of an imprecise basis. The use of other methods
  such as the Jacobi-Davidson iteration to create the basis is
  currently being investigated.

\item For a given blocking, create a locally orthogonal basis using
  the results of step 1. This defines the mapping between coarse and
  fine grid.

\item Solve a second round of Chebyshev-accelerated IRL on the coarse
  Grid to obtain the full set of eigenvectors.

\item Reconstruct an approximation of the eigenvalues by locally
  inverting the Chebyshev polynomial of the Lanczos eigenvalues.

\item The first eigenvalues outside of the basis $N + 1,N + 2,\ldots$ may
  lack sufficient precision which we correct by smoothening the
  corresponding eigenvectors (currently with low-iteration CG) and
  then determining the precise fine-grid eigenvalues.
\end{enumerate}

\begin{figure}[tbp]
  \centering
  \includegraphics[width=9cm]{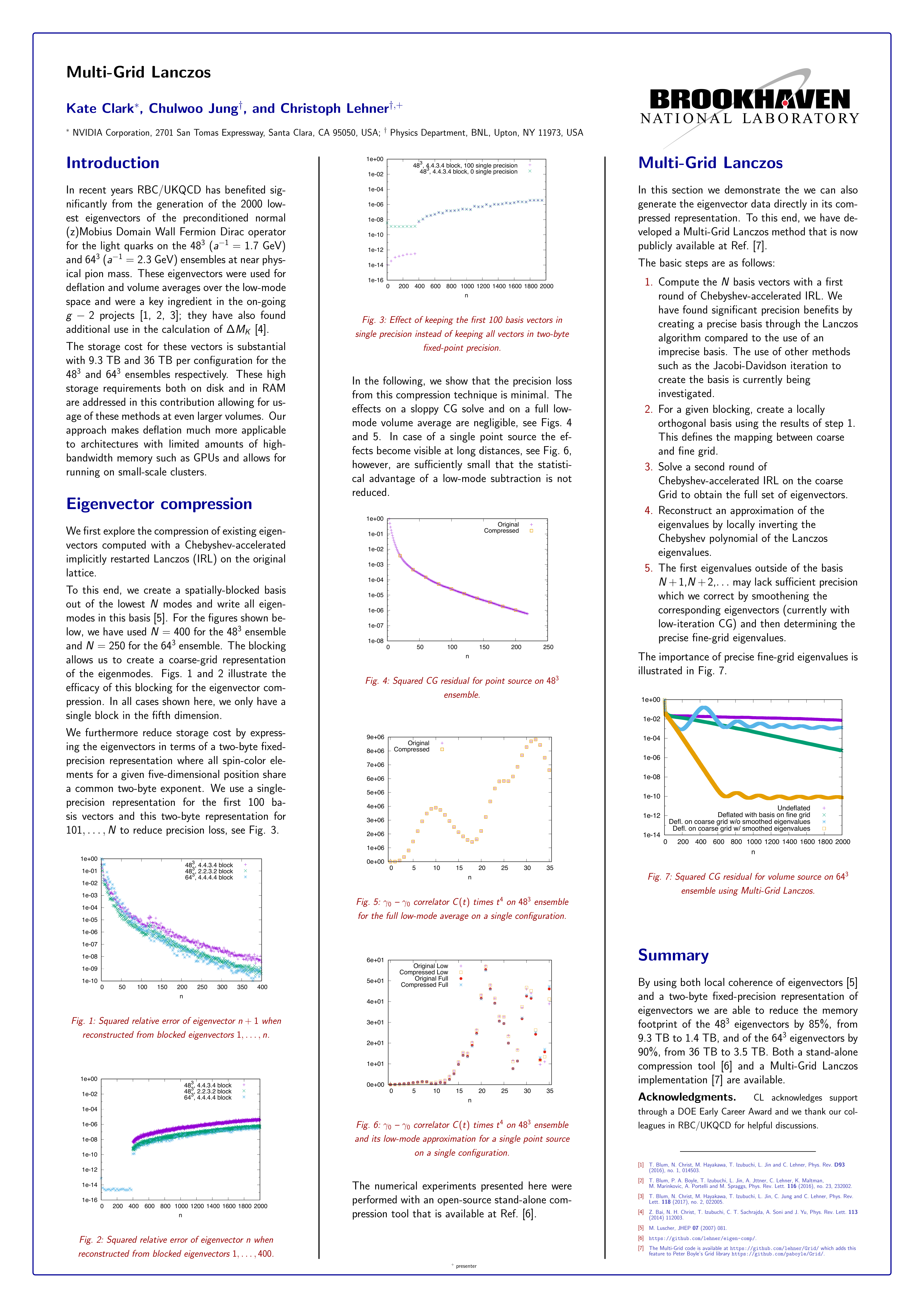}
  \caption{Squared CG residual as function of iteration number for volume source on 64I ensemble using Multi-Grid Lanczos.}
  \label{fig7}
\end{figure}

The importance of precise fine-grid eigenvalues is illustrated in Fig.~\ref{fig7}.

\section{Summary}
By using both local coherence of eigenvectors \cite{ref5} and a
two-byte fixed-precision representation of eigenvectors we are able to
reduce the memory footprint of the 48I eigenvectors by $85\%$, from
9.3 TB to 1.4 TB, and of the 64I eigenvectors by $90\%$, from 36 TB
to 3.5 TB. Both a stand-alone compression tool \cite{ref6} and a
Multi-Grid Lanczos implementation \cite{ref7} are available.

{\bf Acknowledgments.}
C.L.~acknowledges support through a DOE Office of Science Early
Career Award and by US DOE Contract DESC0012704(BNL).
We thank our colleagues in RBC/UKQCD for helpful discussions.

\clearpage
\bibliography{lattice2017}

%%%%%%%%%%%%%%%%%%%%%%%%%%%%%%%%%%%%%%%%%%%%%%%%%%%%%%%%%%%%%%%%%%%%%%%%%%%%%
\end{document}